\documentclass[conference]{IEEEtran}   	

\IEEEoverridecommandlockouts 

\usepackage[utf8]{inputenc}
\usepackage{amsmath,amssymb,amsfonts,amsthm}
\usepackage[numbers,sort&compress]{natbib}

\usepackage{bm}
\usepackage{color}
\usepackage{glossaries}
\usepackage{graphicx}
\usepackage{multirow}
\usepackage[linesnumbered]{algorithm2e}
\usepackage{algpseudocode}
\usepackage{url}

\newacronym{BS}{BS}{base station}
\newacronym{DAC}{DAC}{digital-to-analog converter}
\newacronym{DMA}{DMA}{dynamic metasurface antenna}
\newacronym{FD}{FD}{fully-digital}
\newacronym{KKT}{KKT}{Karush-Khun-Tucker}
\newacronym{MIMO}{MIMO}{multiple-input multiple-output}
\newacronym{MSE}{MSE}{mean squared error}
\newacronym{RF}{RF}{radio-frequency}
\newacronym{5G}{5G}{fifth generation}

\renewcommand{\exp}[1]{\text{exp}\left(#1\right)} 
 
\newcommand{\tr}[1]{\text{Tr}\left\{#1\right\}}	
\renewcommand{\vec}[1]{\mathbf{\lowercase{#1}}}	   
\newcommand{\mat}[1]{\mathbf{\uppercase{#1}}}	   
\newcommand{\real}[1]{\text{Re}\left\{#1\right\}}	

\newcommand{\imag}[1]{\text{Im}\left\{#1\right\}}	
\newcommand{\der}[2]{\frac{\partial \,#1}{\partial \,#2}}  

\definecolor{dkgreen}{rgb}{0.4, 0.7, 0.2}

\title{
Energy efficiency of DMAs vs. conventional MIMO: a sensitivity analysis\\
\thanks{This work has been funded by the European Union under the Marie Sklodowska-Curie grant agreement No. 101109529. The work of D. Morales-Jimenez is supported in part by the State Research Agency (AEI) of Spain and the European Social Fund under grant RYC2020-030536-I and by MICIU/AEI/10.13039/501100011033 and FEDER/UE under grant PID2023-149975OB-I00 (COSTUME).}
\thanks{This work has been submitted to the IEEE for publication. Copyright may be transferred without notice, after which this version may no longer be accesible.}}

\author{Pablo Ram\'irez-Espinosa\IEEEauthorrefmark{1}, David Morales-Jim\'enez\IEEEauthorrefmark{2} and Beatriz Soret\IEEEauthorrefmark{1}\\
\IEEEauthorblockA{\IEEEauthorrefmark{1}Telecommunications Research Institute (TELMA), University of M\'alaga, M\'alaga 29071 (Spain)}
\IEEEauthorblockA{\IEEEauthorrefmark{2}Department of Signal Theory, Networking and Communications, University of Granada, Granada 18071 (Spain)\\
Email: pre@ic.uma.es, dmorales@ugr.es and bsoret@ic.uma.es
}}

\begin{document}
\maketitle

\begin{abstract}
    Motivated by the stringent and challenging need for `greener communications' in increasingly power-hungry 5G networks, this paper presents a detailed energy efficiency analysis for three different multi-antenna architectures, namely fully-digital arrays, hybrid arrays, and dynamic metasurface antennas (DMAs). By leveraging a circuital model, which captures mutual coupling, insertion losses, propagation through the waveguides in DMAs and other electromagnetic phenomena, we design a transmit Wiener filter solution for the three systems. We then use these results to analyze the energy efficiency, considering different consumption models and supplied power, and with particular focus on the impact of the physical phenomena. 
    DMAs emerge as an efficient alternative to classical arrays across diverse tested scenarios, most notably under low transmission power, strong coupling, and scalability requirements.  
\end{abstract}

\section{Introduction}
\label{sec:intro}
\Gls{5G} networks have proved to be more power hungry than previous technologies despite achieving the largest efficiency in terms of transmitted bits per Joule \cite{GSMA2022}. Most of this power (up to 73$\%$) is consumed at \glspl{BS}, with power amplifiers being the most demanding components, followed by the \gls{RF} chains and the computation at the base-band unit \cite{Ge2017, GSMA2021}. The trend towards denser \gls{BS} infrastructures, the adoption of massive arrays and/or higher frequency bands (with reduced hardware efficiency) further increases this consumption. 

Therefore, there is a growing need to look for \textit{greener} hardware implementations at \glspl{BS}, not only by improving the efficiency of \gls{RF} devices, but also by reducing their number and 
replacing them by more efficient solutions. 
Hybrid \gls{MIMO} is a classical alternative to \gls{FD} topologies, where a complete \gls{RF} chain and amplifier are required per each antenna. Recently, the scientific community has turned to \glspl{DMA} as a cheaper and more efficient alternative to hybrid topologies \cite{You2023, Carlson2024, Williams2022}, in which the phased-arrays are replaced by radiating elements controlled by semiconductor devices. 

Although apparently inferior to hybrid \gls{MIMO} in terms of raw performance \cite{Zhang2022, Ramirez2022}, the energy efficiency of these architectures has not been properly characterized yet. To the best of our knowledge, only a couple of works tackle the problem \cite{Carlson2024, You2023}, reporting a significantly higher efficiency of \glspl{DMA} with respect to classical arrays. In \cite{Carlson2024}, a simplistic model for the \gls{DMA} is assumed, ignoring insertion losses and mutual coupling \cite{Ramirez2022}. In \cite{You2023}, a more realistic full-wave simulated \gls{DMA} is considered, but the analysis is restricted to a single user case and codebook-based precoding, which  opens the question of the achievable energy efficiency in the multi-user case. Besides, power consumption studies are heavily dependent on the selected model, with results varying significantly based on factors such as device implementation and frequency band. Thus, a performance evaluation under different modeling options is advisable.

In this work, we carry out a thorough evaluation of the energy efficiency achieved by conventional \gls{FD} and hybrid arrays, as well as by \glspl{DMA} as a contender solution. The three topologies are characterized using circuit theory, inheriting the model in \cite{Williams2022} and capturing electromagnetic phenomena such as mutual coupling and insertion losses; these aspects are usually overlooked in the literature. Besides, a transmit Wiener filter precoder is proposed for the three architectures. Interestingly, we show that \glspl{DMA} arise as promising and efficient solutions, specially when lower transmit powers are available and in cases where we need to pack a massive number of antennas (radiating elements) in constrained apertures.


\textit{Notation:} Vectors and matrices are represented by bold lowercase and uppercase symbols, respectively. $(\cdot)^T$ and $(\cdot)^H$ denote transpose and conjugate transpose, $\tr{\cdot}$ is the trace, and $\|\cdot\|_2$ is the $\ell_2$ norm of a vector. Also, $\mat{I}_n$ is the identity matrix of size $n\times n$, and $(\mat{A})_{j,k}$ is the $j,k$-th element of $\mat{A}$. Finally, $i = \sqrt{-1}$ is the imaginary number, $\mathbb{E}_{x}[\cdot]$ is the mathematical expectation over $x$, $\circ$ denotes the Hadamard (element-wise) product, and $\real{\cdot}$ and $\imag{\cdot}$ denote real and imaginary part. 

\section{Communication Model for MIMO systems}
\label{sec:Commun_Model}
\begin{figure*}[t]
    \centering
    \includegraphics[width=0.8\linewidth]{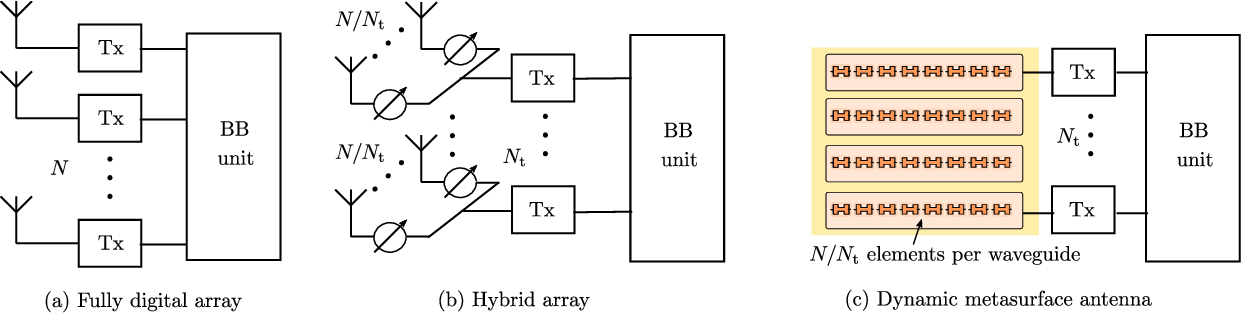}
    \caption{MIMO topologies. Each transmitter (Tx) is composed by a DAC, an RF chain and a power amplifier, and is directly fed by the base-band unit.}
    \label{fig:MIMO_topologies}
\end{figure*}

We consider a generic cellular communications system, where a \gls{BS} serves $M$ users simultaneously. The \gls{BS} is equipped with $N_\text{t}$ transmitters, each one composed by a pair of \glspl{DAC}, an \gls{RF} chain, and a power amplifier. Each transmitter feeds either a single antenna element in an \gls{FD} array, a tunable phased array (partially connected hybrid array), or a single waveguide in a \gls{DMA}, as illustrated in Fig. \ref{fig:MIMO_topologies}. Regardless of the chosen topology, $N$ denotes the total number of antennas or radiating elements, and the $M$ users are equipped with single-antenna devices.

The system is analyzed by using circuit theory, inheriting the models for \glspl{DMA} and arrays from \cite{Williams2022, Ramirez2022}, while generalizing the latter to account also for insertion losses. Every antenna is modeled as a magnetic dipole, located at an arbitrary position  $\vec{r}_n = (x_n, y_n, z_n)^T$, and oriented along the $z$ direction. This type of theoretical antenna characterizes well the radiating properties of the elements used in \glspl{DMA} \cite{Smith2017} without loss of generality in the analysis of conventional arrays. The \gls{DMA} or the array at the \gls{BS} is located in the $xz$-plane, while the users are distributed in the half-space defined by $y>0$. 

\subsection{Circuital analysis of conventional arrays}

Denoting by $\vec{j}_\text{r}\in\mathbb{C}^{M\times 1}$ and by $\vec{j}_\text{t}\in\mathbb{C}^{N\times 1}$ the magnetic currents at the users' loads and at the \gls{BS} antennas, respectively, in the case of \gls{FD} and hybrid \gls{MIMO} we have \cite{Williams2022, Ramirez2022} 
\begin{equation}
    \vec{j}_\text{r} = -\frac{1}{2}\mat{Y}_\text{r}^{-1}\mat{Y}_\text{ra}\vec{j}_\text{t}, \label{eq:jr_FD_jt}
\end{equation}
where $\mat{Y}_\text{r}=Y_\text{r}\mat{I}_M$ is a diagonal matrix containing the (real) users' load admittances (note that conjugate loads are assumed at the users) and $\mat{Y}_\text{ra}\in\mathbb{C}^{M\times N}$ is the wireless channel matrix. Modeling each transmitter (or each phase shifter in the hybrid topology) as a Norton's equivalent generator, the relation 
\begin{equation}
    \vec{j}_\text{t} = 2Y_\text{g}(Y_\text{g}\mat{I}_N + \mat{Y}_\text{aa})^{-1}\vec{j}_\text{g} \label{eq:Relation_jt_jg}
\end{equation} 
is obtained (it can be easily proved from \cite[Sec. III-A]{Williams2022}), where $Y_\text{g}\in\mathbb{R}^+$ is the intrinsic admittance of the generator, $\vec{j}_\text{g}\in\mathbb{C}^{N\times 1}$ is the supplied currents vector and $\mat{Y}_\text{aa}\in\mathbb{C}^{N\times N}$  is the coupling matrix of the array. Assuming the array is on a conductor plane, $\mat{Y}_\text{aa}$ is given by 
\begin{align}
    (\mat{Y}_\text{aa})_{n,n'} =&  \begin{cases}
    i2\omega \epsilon G_a\left(\vec{r}_n, \vec{r}_{n'}\right)  & n \neq n' \\   k\omega \epsilon/{(3\pi)}        & n = n'
    \end{cases}, \label{eq:Yq_fd} 
\end{align}
with $\omega$ the angular frequency, $\epsilon$ the electrical permittivity, $k=2\pi/\lambda$ the wavenumber and $G_a(\cdot)$ as in \cite[Eq. (39)]{Williams2022}. Finally, the supplied power to the array is calculated as
\begin{equation}
    P_g = \frac{1}{2}\mathbb{E}\left[\real{Y_\text{g}\vec{j}_\text{g}^H\vec{j}_\text{g}}\right]. \label{eq:Pg_currents}
\end{equation}

\subsection{Circuital analysis of DMAs}
\label{sec:CircuitDMA}

We consider the \gls{DMA} structure illustrated in Fig. \ref{fig:MIMO_topologies}, where multiple waveguides are stacked in parallel with radiating elements attached on top of them and controlled through varactor diodes. In this case, the induced currents at users' loads are given by
\begin{equation}
    \vec{j}_\text{r} = \frac{1}{2}\mat{Y}_\text{r}^{-1}\mat{Y}_\text{rs}(\mat{Y}_\text{s}+\mat{Y}_\text{ss})^{-1}\mat{Y}_\text{st}\vec{j}_\text{t}, \label{eq:jr_jt_dma}
\end{equation}
where $\vec{j}_\text{t}\in\mathbb{C}^{N_\text{t}\times 1}$ represent, in this case, the currents entering the waveguides (and not the radiating elements), $\mat{Y}_\text{st}\in\mathbb{C}^{N\times N_\text{t}}$ is the mutual admittance between transmitters and the radiating elements on each waveguide, $\mat{Y}_\text{ss}\in\mathbb{C}^{N\times N}$ is the coupling matrix between the elements, and $\mat{Y}_\text{rs}\in\mathbb{C}^{M\times N}$ is the wireless channel ($\mat{Y}_\text{rs} = \mat{Y}_\text{ra}$ if the antennas are placed at the same positions in all the topologies). The reconfigurability of the structure is encapsulated in the load admittance $\mat{Y}_\text{s} = R_\text{s}\mat{I}_N + i\mat{Y}_\text{s}^\text{im}$, with $R_\text{s}\in\mathbb{R}^+$ representing the antenna losses and $\mat{Y}_\text{s}^\text{im}\in\mathbb{R}^{N\times N}$ a tunable diagonal matrix controlled by the varactor diodes. As with the arrays, we can write 
\begin{equation}
    \vec{j}_\text{t} = 2Y_\text{g}(Y_\text{g}\mat{I}_N + \mat{Y}_\text{p})^{-1}\vec{j}_\text{g}, \label{eq:Relation_jt_jg_dma}
\end{equation} 
where $\mat{Y}_\text{p} = \mat{Y}_\text{tt} - \mat{Y}_\text{st}^T(\mat{Y}_\text{s}+\mat{Y}_\text{ss})^{-1}\mat{Y}_\text{st}$, and the supplied power to the whole structure is given by \eqref{eq:Pg_currents}. Closed-form expressions for $\mat{Y}_\text{st}$, $\mat{Y}_\text{ss}$ and $\mat{Y}_\text{tt}$ are provided in \cite[Table 1]{Williams2022}. 

\subsection{MIMO signal model}
Following standard \gls{MIMO} signal models, we express the received signal at the users, $\vec{y}\in\mathbb{C}^{M\times 1}$, as
\begin{equation}
    \vec{y} = \mat{H}_\text{eq}\mat{B}\vec{x} + \vec{n}, \label{eq:y_general}
\end{equation}
where $\mat{H}_\text{eq}$ is the equivalent channel, $\mat{B}$ is some precoding matrix, $\vec{x}\in\mathbb{C}^{M\times 1}$ is the intended message for the users and $\vec{n}\in\mathbb{C}^{M\times 1}$ is the noise term. As general considerations, we assume $\mathbb{E}[\vec{x}\vec{x}^H] = \mat{I}_M$, $\vec{n}\sim\mathcal{CN}_M(\vec{0},\sigma_n^2\mat{I}_M)$, and statistical independence between $\vec{x}$ and $\vec{n}$. Since \eqref{eq:y_general} relates transmitted and received signal (in terms of baseband voltages), it is natural to relate $\vec{j}_\text{r}=\mat{H}_\text{eq}\mat{B}\vec{x}$ (noise-free component of $\vec{y}$) and $\vec{j}_\text{g} = \mat{B}\vec{x}$, since $\vec{j}_\text{g}$ is the output of the transmitters and thus the signal we can control, letting $\mat{H}_\text{eq}$ to capture all the circuital characteristics of each topology. Next, we particularize \eqref{eq:y_general} for each system.


\subsubsection{Fully digital and hybrid systems}

From \eqref{eq:jr_FD_jt}-\eqref{eq:Relation_jt_jg}, we get
\begin{equation}
    \mat{H}_\text{eq}\rvert_\text{array} = \mat{H}_\text{a} = -\alpha_r Y_\text{g}\mat{Y}_\text{r}^{-1}\mat{Y}_\text{ra}(Y_\text{g}\mat{I}_N+\mat{Y}_\text{aa})^{-1}, \label{eq:H_a}
\end{equation}
where $\alpha_r = \sqrt{\real{Y_\text{r}}/2}$ is a scaling factor such that $\mathbb{E}[\|\vec{y}\rvert_{\vec{n}=\vec{0}}\|_2^2] = \frac{1}{2}\mathbb{E}[\real{Y_\text{r}\vec{j}_\text{r}^H\vec{j}_\text{r}}]$ (averaged received power). In the case of an \gls{FD} array, then $\mat{B}\in\mathbb{C}^{N\times M}$ is a digital precoder, while $\mat{B} = \mat{Q}\mat{B}_\text{h}$ in the hybrid case, where $\mat{B}_\text{h}\in\mathbb{C}^{N_\text{t}\times M}$ is the digital precoder and $\mat{Q}\in\mathbb{C}^{N\times N_\text{t}}$ is the phase shifting matrix.  Considering a lossless and perfectly matched phase shifting network, $(\mat{Q})_{n,n'} = e^{i\theta_{n,n'}}$ with $\theta_{n,n'}\in\mathbb{R}$ if antenna $n$ is connected to the $n'$-th transmitter and $(\mat{Q})_{n,n'} = 0$ otherwise.
Using the relation $\vec{j}_\text{g}=\mat{B}\vec{x}$, the supplied power is simply \eqref{eq:Pg_currents}
\begin{equation}
    P_g = \frac{1}{2}{Y_\text{g}}\tr{\mat{B}^H\mat{B}}. \label{eq:Pg_MIMO_array}
\end{equation}

\subsubsection{DMA}

From \eqref{eq:jr_jt_dma} and \eqref{eq:Relation_jt_jg_dma}, we have $\mat{H}_\text{eq}\rvert_\text{dma} = \mat{H}_\text{d}$ with
\begin{equation}
     \mat{H}_\text{d} = \alpha_r Y_\text{g}\mat{Y}_\text{r}^{-1}\mat{Y}_\text{rs}(\mat{Y}_\text{s}+\mat{Y}_\text{ss})^{-1}\mat{Y}_\text{st}(Y_\text{g}\mat{I}_{N_\text{t}} + \mat{Y}_\text{p})^{-1}. \label{eq:Heq_dma}
\end{equation}
Similarly, $\mat{B}\in\mathbb{C}^{N_\text{t}\times M}$ is a digital precoder (applied at the transmitters) and the supplied power is directly given by \eqref{eq:Pg_MIMO_array}. Finally, note from \cite[Sec. V-A]{Williams2022} that $\alpha_r\mat{Y}_\text{r}^{-1} = \sqrt{\frac{3\pi}{k\omega\epsilon}}\mat{I}_M = \widetilde{\alpha}_\text{r}\mat{I}_M$.

\section{Power Consumption modelling}
\label{sec:Power_Model}

Characterizing the power consumption of a \gls{MIMO} system is by no means a trivial task, since it heavily depends on the specific implementation and hardware. Our goal is to provide a sufficiently versatile and general (albeit simple) consumption model based on previous work \cite{Auer2011, Carlson2024, You2023, Ribeiro2018, Mendez2016} that allows us to compare the three architectures as fairly as possible. 

In general, the power consumption of each architecture is given by the sum of all its components' consumption. Thus, we define the total power consumption as
\begin{equation}
    P_\text{total} = P_\text{bb} + N_\text{t}\left(2P_\text{dac} + P_\text{rf}\right) + P_\text{a} + N_\text{ps}P_\text{ps} + N_\text{var}P_\text{var},
\end{equation}
where each of the terms are explained next. First, $P_\text{bb}$ is the consumption at the base-band unit, which computes the precoding and the control signals for the phase-shifting network and the \gls{DMA}. $P_\text{dac}$ is the power consumption at the \glspl{DAC}, which is accurately modelled as \cite[Eq. (13)]{Ribeiro2018}
\begin{equation}
    P_\text{dac} = 1.5\cdot10^{-5}2^b + 9\cdot10^{-12}b F_s,
\end{equation}
where $b$ is the number of resolution bits and $F_s$ the maximum sampling frequency. $P_\text{rf}$ refers to the power consumption at the \gls{RF} chains, which usually involves small-signal analog operations such as up-conversion, filtering and some amplification. Its consumption heavily depends on the specific implementation, ranging from a few milliwatts for small cells \cite{Mendez2016} to several watts in macrocells \cite{Auer2011}. $P_{\text{a}}$ is the consumption at the power amplifiers, for which two modelling choices are proposed in the literature: \text{i)} a linear model where $P_\text{a} = P_g/\eta_a$, i.e., the ratio of the output power $P_g$ and the amplifier efficiency $\eta_a$; and \textit{ii)} a non-linear model where $P_{\text{a}} = \sum_n\sqrt{P_{\text{out},n}P_\text{sat}}/\eta_a$ with $P_\text{sat}$ the saturation output power and $P_{\text{out},n}$ the power delivered by the amplifier \cite{Persson2013}. The latter model is more realistic, since power amplifiers are more efficient as they approach their maximum output power, although being detrimental for communication purposes. Note also that $\sum P_{\text{out},n} = P_g$.

Besides, $N_\text{ps}$ and $P_\text{ps}$ characterize, respectively, the number of phased shifters in the hybrid system and their power consumption ($N_\text{ps}=0$ for \gls{FD} and \gls{DMA} based systems). Active phase shifters are generally preferred over passive ones due to their lower insertion losses and smaller footprint \cite{Ribeiro2018, Mendez2016}. Similarly, $N_\text{var}$ and $P_\text{var}$ are the number and consumption of varactor diodes in \glspl{DMA} ($N_\text{var}=0$ in conventional arrays). However, this consumption has proved to be negligible \cite{Ribeiro2018}, i.e., $P_\text{var}\approx 0$. Therefore, the main difference between hybrid arrays and \glspl{DMA} in terms of energy efficiency is brought by the impact of $P_\text{ps}$. Please note that, for simplicity, the feeding network connecting the transmitters to the different radiating elements, as well as the waveguides in the \gls{DMA} structure, are assumed to be lossless.

\section{Precoding Design: Transmit Wiener Filter}
\label{sec:Precoding}
The Wiener filter---equivalently, unweighted minimum \gls{MSE}---is obtained by minimizing the \gls{MSE} between the transmitted symbols $\vec{x}$ and the scalar-weighted received signal under some power constraint \cite[Eq. (36)]{Joham2005}. In the following, we provide a solution for the three considered architectures, where the supplied power by the transmitters is used as constraint and perfect channel information is assumed. 

\subsection{Solution for FD arrays}
In this case, the optimization problem is formulated as
\begin{subequations} \label{eq:WF_fd}
\begin{align} 
~~ \mathop {\text{minimize}}\limits_{\mat{B},\;\beta} \;\;&\mathbb{E}_{\vec{x},\vec{n}}\left[\|\vec{x} - \beta^{-1}(\mat{H}_\text{a}\mat{B}\vec{x}+\vec{n})\|_2^2\right]\hfill \\
{\text{s.t.}}\;& \frac{1}{2}Y_\text{g}\tr{\mat{B}^H\mat{B}} \leq P_g^\text{max}, \label{eq:Pconst_FD}\hfill
\end{align}
\end{subequations}
where $\beta\in\mathbb{R}^+$ and $P_g^\text{max}\in\mathbb{R}^+$ is the maximum supplied power. 
As proved in \cite{Joham2005}, the above problem has a closed-form solution when the weight $\beta$ is positive and real, representing a power scaling carried out at the users (a diagonal receive filter with constant entries). Hence, solving the \gls{KKT} conditions, we obtain\footnote{The proof is omitted due to space constraints.} $\mat{B} = \beta\mat{A}^{-1}\mat{H}_\text{a}^H$ with
    \begin{align}
    \mat{A} &=  \left(\mat{H}_\text{a}^H\mat{H}_\text{a} + \tfrac{M\sigma_n^2}{2P_g^\text{max}}Y_\text{g}\mat{I}_N\right)^{-1}\label{eq:A_FD} \\
    \beta &= \sqrt{\frac{2P_g^\text{max}}{Y_\text{g}\text{Tr}\left\{\mat{H}_\text{a}\mat{A}^{-2}\mat{H}_\text{a}^H\right\}}}.
\end{align}

\subsection{Solution for hybrid arrays}

Analogously to the \gls{FD} case, the problem is formulated as
\begin{subequations}\label{eq:WF_Hybrid}
\begin{align} 
~~ \mathop {\text{minimize}}\limits_{\mat{Q},\mat{B}_\text{h},\beta} \;\;&\mathbb{E}_{\vec{x},\vec{n}}\left[\|\vec{x} - \beta^{-1}(\mat{H}_\text{a}\mat{Q}\mat{B}_\text{h}\vec{x}+\vec{n}))\|_2^2\right]\hfill \\
{\text{s.t.}}\;& \frac{1}{2}Y_\text{g}\tr{\mat{B}_\text{h}^H\mat{Q}^H\mat{Q}\mat{B}_\text{h}} \leq P_g^\text{max}, \hfill \\
& |(\mat{Q})_{n,n'}| = 1 \quad\forall\quad (\mat{Q})_{n,n'} \neq 0.\hfill \label{eq:Q_const_Hybrid}
\end{align}
\end{subequations}

To the best of our knowledge, no closed-form solution is known for \eqref{eq:WF_Hybrid} due to the unit modulus constraint, and similar problems are usually tackled by optimizing $\mat{B}_\text{h}$ and $\mat{Q}$ separately. For fixed $\mat{Q}$, \eqref{eq:WF_Hybrid} is very similar to \eqref{eq:WF_fd}, and the optimal solutions for $\beta$ and $\mat{B}_\text{h}$ are thus obtained by solving the \gls{KKT} conditions, yielding
\begin{align}
   \mat{B}_\text{h} &= \beta\left(\mat{Q}^H\mat{A}\mat{Q}\right)^{-1}\mat{Q}^H\mat{H}_\text{a}^H, \label{eq:B_Hybrid_WF} \\
   \beta &= \sqrt{\frac{2P_g^\text{max}}{Y_\text{g}\tr{\mat{H}_\text{a}\left(\mat{Q}(\mat{Q}^H\mat{A}\mat{Q})^{-1}\mat{Q}^H\right)^{2}\mat{H}_\text{a}^H}}},
\end{align}
with $\mat{A}$ as in \eqref{eq:A_FD}. To derive $\mat{Q}$, we introduce the above results into the objective function in \eqref{eq:WF_Hybrid}, leading to
\begin{equation}
    \text{MSE}_\text{h} = \tr{\mat{I}_M-\mat{H}_\text{a}\mat{Q}(\mat{Q}^H\mat{A}\mat{Q})^{-1}\mat{Q}^H\mat{H}_\text{a}^H}, \label{eq:MSE_hybrid_full}
\end{equation}
and the optimum matrix $\mat{Q}$ is the one that minimizes the above \gls{MSE}. Since $\mat{Q}$ can be expressed as $\mat{Q} = \exp{i\bm{\Theta}}\circ\mat{S}$, where $\mat{S}$ is a selection matrix indicating the non-zero elements of $\mat{Q}$, we can directly optimize over $\bm{\Theta}$, removing the constraint \eqref{eq:Q_const_Hybrid}, i.e., solve $\min_{\bm{\Theta}} \;\;\text{MSE}_\text{h}$, whose gradient is given by
\begin{align}
    \der{\text{MSE}_\text{h}}{\bm{\Theta}} = 2\imag{\left(\mat{C}_h\mat{H}_\text{a}^H\mat{H}_\text{a}\left[\mat{I}_N-\mat{Q}\mat{C}_h\mat{A}\right]\right)^T\circ\mat{Q}}, \label{eq:Gradient_Hybrid}
\end{align}
where, for the sake of notation, we have defined $\mat{C}_h = \left(\mat{Q}^H\mat{A}\mat{Q}\right)^{-1}\mat{Q}^H$. Due to space constraints, the proof of \eqref{eq:Gradient_Hybrid} is omitted, and the reader is referred to \cite{Hjrungnes2011} for standard methods on matrix derivatives. From \eqref{eq:Gradient_Hybrid}, numerical methods like gradient descent or quasi-Newton algorithms suffice to find a solution.

\subsection{Solution for DMAs}

In this case, the Wiener filter is the solution to
\begin{subequations}\label{eq:WF_DMA}
\begin{align} 
~~ \mathop {\text{minimize}}\limits_{\mat{B},\;\beta,\;\mat{Y}_\text{s}} \;\;&\mathbb{E}_{\vec{x},\vec{n}}\left[\|\vec{x} - \beta^{-1}(\mat{H}_\text{d}\mat{B}\vec{x}+\vec{n}))\|_2^2\right]\hfill \\
{\text{s.t.}}\;& \frac{1}{2}Y_\text{g}\tr{\mat{B}^H\mat{B}} \leq P_g^\text{max}, \label{eq:Pconst_DMA}\hfill
\end{align}
\end{subequations}
where a direct solution is difficult---if not impossible---due to the complicated dependence on $\mat{Y}_\text{s}$. Following the same steps as previously, we assume first fixed $\mat{Y}_\text{s}$ (and hence fixed $\mat{H}_\text{d}$), leading to the closed form solution $\mat{B} = \beta\widetilde{\mat{A}}^{-1}\mat{H}_\text{d}^H$ with
\begin{align} 
    \widetilde{\mat{A}} &= \left(\mat{H}_\text{d}^H\mat{H}_\text{d} + \tfrac{M\sigma_n^2}{2P_g^\text{max}}Y_\text{g}\mat{I}_{N_\text{t}}\right)^{-1}, \label{eq:B_DMA_WF} \\
    \beta_\text{d} &= \sqrt{\frac{2P_g^\text{max}}{Y_\text{g}\tr{\mat{H}_\text{d}\widetilde{\mat{A}}^{-2}\mat{H}_\text{d}^H}}}.
\end{align}

Introducing this solution into the objective function in \eqref{eq:WF_DMA} and using the matrix inversion lemma yield $\text{MSE}_\text{d} = \tr{\mat{G}^{-1}}$ where
\begin{equation}
    \mat{G} = \mat{I}_M + \tfrac{2P_g^\text{max}}{M\sigma_n^2Y_\text{g}}\mat{H}_\text{d}\mat{H}_\text{d}^H, \label{eq:MSE_dma}
\end{equation}
and the optimum $\mat{Y}_\text{s}$ is obtained by minimizing it. Since only the imaginary part of $\mat{Y}_\text{s}$ can be controlled, as stated in Section \ref{sec:CircuitDMA}, we directly optimize over $\mat{Y}_\text{s}^\text{im}$, aiming to solve $\min_{\mat{Y}_\text{s}^\text{im}} \text{MSE}_\text{d}$,
which is unconstrained with gradient given by \eqref{eq:GradientDMA}, placed at the top of the next page. Due to space limitations, the proof is also omitted.

\begin{figure*}[t]
\begin{align}
    \der{\text{MSE}_\text{d}}{\mat{Y}_\text{s}^\text{im}} = -\tfrac{4P_g^\text{max}\widetilde{\alpha}_\text{r}}{M\sigma_n^2}\imag{\left[(\mat{Y}_\text{s}+\mat{Y}_\text{ss})^{-1}\mat{Y}_\text{st}(\mat{Y}_\text{p}+Y_\text{g}\mat{I}_{N_\text{rf}})^{-1}\mat{H}_\text{d}^H\mat{G}^{-2}\left[\mat{Y}_\text{rs}+\tfrac{1}{Y_\text{g}\widetilde{\alpha}_\text{r}}\mat{H}_\text{d}\mat{Y}_\text{st}^T\right](\mat{Y}_\text{s}+\mat{Y}_\text{ss})^{-1}\right]\circ\mat{I}_N}. \label{eq:GradientDMA}
\end{align}
\hrulefill
\end{figure*}
\section{Performance Evaluation and Discussion}
\label{sec:Evaluation}

\begin{table}[t]
    \renewcommand{\arraystretch}{1.2}
    \centering
    \caption{Baseline set of simulation parameters}
    \begin{tabular}{c|c||c|c}
         \textbf{Parameter} & \textbf{Value} & \textbf{Parameter} & \textbf{Value}  \\ \hline\hline
         $\sigma_n^2$ & $0.02$ & $P_\text{bb}$ & $40$ mW \\ \hline
         $Y_\text{g}$ (array) & $k\omega\epsilon/(3\pi)$ & $P_\text{dac}$ & $11$ mW \\ \hline
         $Y_\text{g}$ (DMA) & $35.33$ & $P_\text{rf}$ & $40$ mW \\ \hline
         $a$ & $0.73\lambda$ & $P_\text{ps}$ & $21.8$ mW \\ \hline
         $b$ & $0.17\lambda$ & $P_\text{var}$ & $0$ mW \\ \hline
         $L_w$ & s.t. $k_xL_w = 3\pi/4$ & $\eta_a$ & $0.3$  \\ \hline
         frequency & $10$ GHz & $P_\text{sat}\rvert_\text{fd}$, $P_\text{sat}\rvert_\text{other}$ & $\frac{5P_g^\text{max}}{N}$, $\frac{3P_g^\text{max}}{N_\text{t}}$ \\
    \end{tabular}
    \label{tab:SimParaBaseline}
\end{table}

Although the three topologies can be tested in a wide variety of scenarios and conditions, we here focus on getting some insight on: \textit{i)} the scaling of \glspl{DMA} and hybrid arrays as we increase the number of antennas per transmitters, \textit{ii)} the impact of the maximum supplied power, \textit{iii)} the sensitivity of the performance to changes in the power consumption model, and \textit{iv)} the impact of mutual coupling and insertion losses. 

For all the simulations, we consider correlated Rayleigh fading for both $\mat{Y}_\text{rs}$ and $\mat{Y}_\text{ra}$, with covariance matrix $\bm{\Sigma} = \widetilde{\alpha}\real{\mat{Y}_\text{aa}}$ (a scaled version of \cite[Eq. (59)]{Williams2022} such that the wireless channels have unit average power). Besides, the noise variance is fixed to a reference value of $\sigma_n^2 = 0.02$, such that $\mat{E}[|(\mat{Y}_\text{ra})_{m,n}|^2]/\sigma_n^2 = 4$ for any $m,n$. Without loss of generality, the frequency is set to $10$ GHz as in \cite[Sec. VII]{Williams2022} (note that the model is written in terms of the wavelength, so varying the frequency only implies scaling the physical dimensions). 
For the array-based architectures, the transmitter self-admittance $Y_\text{g}$ is chosen to achieve matching admittances when there is no mutual coupling, i.e., $Y_\text{g} = k\omega\epsilon / (3\pi)$. For the \gls{DMA}, identical waveguides are considered with dimensions\footnote{$a$ and $b$ are chosen such that only the fundamental mode $\text{TE}_{10}$ propagates.} $a=0.73\lambda$ along the $z$-axis (width), $b=0.17\lambda$ along the $y$-axis (height) and $L_w$ along the $x$-axis (length), where $L_w$ is chosen such that\footnote{$k_x = \sqrt{k^2 -(\pi/a)^2}$.} $k_xL_w = 3\pi/4$ to normalize $\mat{Y}_\text{tt}$ regardless of the \gls{DMA} dimension (c.f. \cite[Eq. (36)]{Williams2022}). $Y_\text{g}$ is chosen in this case to match the waveguide self-admittance, i.e., $Y_\text{g} = 35.33$. 

Regarding the power consumption modeling, unless otherwise stated, we use the baseline values $P_\text{bb} = 40$ mW \cite{Mendez2016}, $P_\text{dac} = 11$ mW ($b = 8$ and $F_s = 100$ MHz), $P_\text{rf} = 40$ mW \cite{Mendez2016}, $P_\text{ps} = 21.8$ mW, $P_\text{var} \approx 0$, and $\eta_a = 0.3$ \cite{Ribeiro2018}. Finally, the non-linear model is assumed for the amplifiers, with $P_\text{sat} = 5P_g^\text{max}/N$ for the \gls{FD} topology and $P_\text{sat} = 3P_g^\text{max}/N_\text{t}$ for the hybrid and \gls{DMA} case. These values have been obtained by simulation, ensuring that the maximum output power of each amplifier\footnote{The output power of $n$-th amplifier is calculated as $Y_\text{g}|(\vec{j}_\text{g})_n|^2/2$ for \gls{FD} and \gls{DMA} and as $Y_\text{g}\frac{N}{N_\text{t}}|(\vec{j}_\text{g})_n|^2/2$ for hybrid arrays.} is at least $3$ dB lower than $P_\text{sat}$. For the reader's sake, this set of parameters is summarized in Table \ref{tab:SimParaBaseline}. 

\begin{figure}[t]
    \centering
    \includegraphics[width=0.8\linewidth, trim={0 0.4cm 0 0.4cm}]{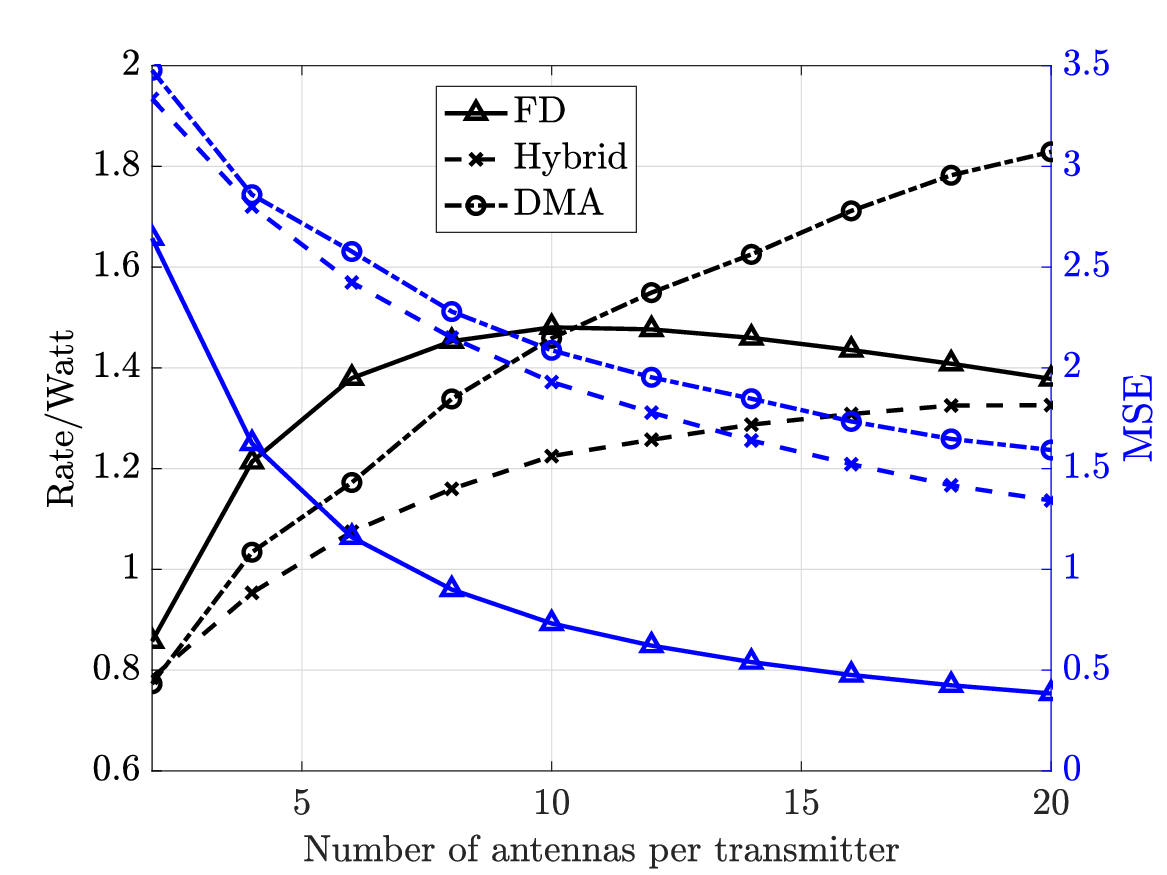}
    \caption{Energy efficiency and average MSE achieved by the three topologies vs $N/N_\text{t}$ for $N_\text{t} = 8$, $M = 6$, $P_g^\text{max} = 30$ dBm, and spacing of $\lambda/2$ along the $x$ axis and $\lambda$ along the $z$ axis.}
    \label{fig:N_evo_ee_mse}
\end{figure}

Under these conditions, we evaluate first in Fig. \ref{fig:N_evo_ee_mse} the performance of the three topologies as we increase the number of antennas per transmitter $N/N_\text{t}$ (in \gls{FD}, naturally, $N=N_\text{t}$ is increased). The mutual coupling is low (though not negligible) for the array topologies, since the spacing is set to $0.5\lambda$ along the $x$ axis (between antennas attached to the same transmitter) and to $\lambda$ along the $z$ axis (separation between waveguides centers in the \gls{DMA}). We see that, despite notably outperforming the other topologies in terms of \gls{MSE} (in blue), the \gls{FD} solution starts decreasing its energy efficiency (in black) after reaching a certain number of antennas. An explanation for this is that, since the number of served users is fixed, increasing the size of the \gls{FD} arrays leads to an ``over-provisioning", considerably increasing the power consumption with diminishing returns. The \gls{DMA} setup, however, always outperforms the hybrid array, and indeed becomes the most efficient solution for large numbers of antennas. 

\begin{figure}[t]
    \centering
    \includegraphics[width=0.8\linewidth, trim={0 0.4cm 0 0.4cm}]{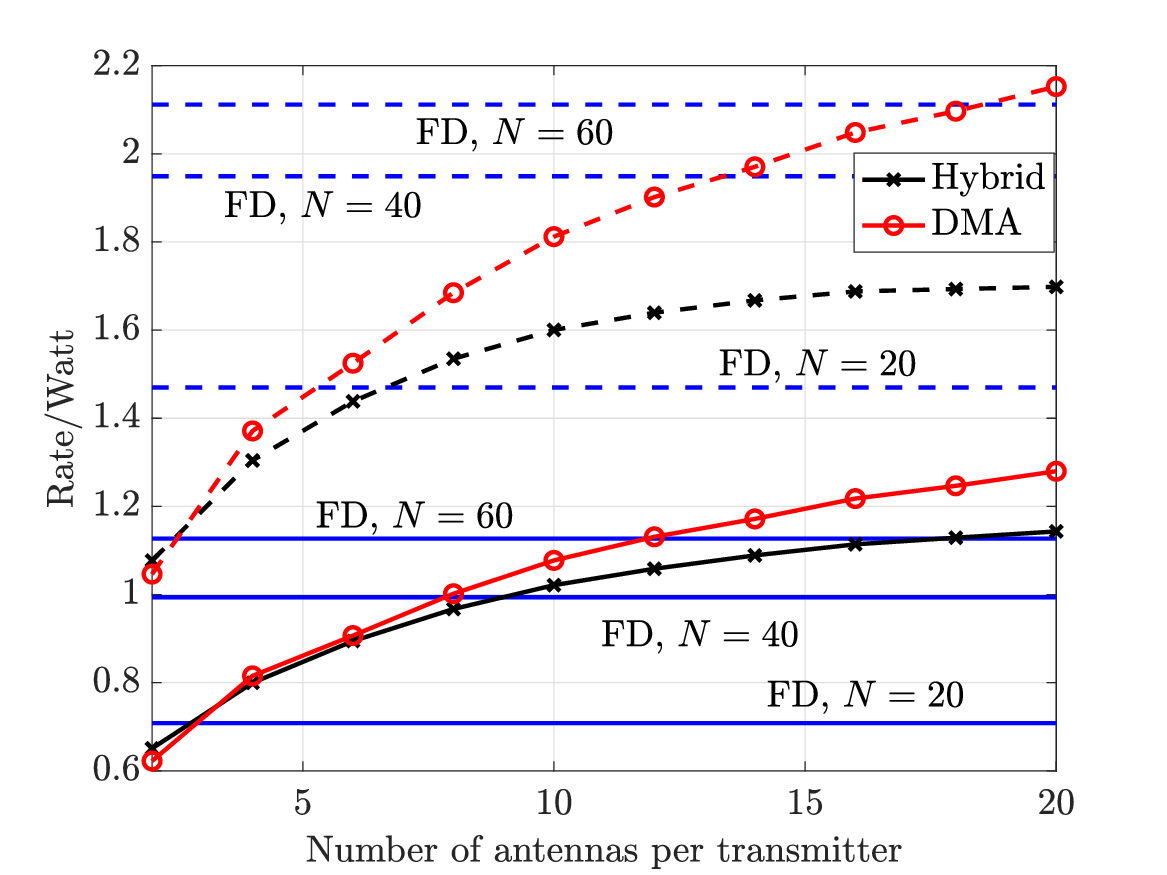}
    \caption{Energy efficiency for hybrid and \gls{DMA} topologies vs $N/N_\text{t}$ for $N_\text{t} = 10$, $M = 8$, $P_g^\text{max} = 45$ dBm, and spacing of $\lambda/2$ along the $x$ axis and $\lambda$ along the $z$ axis. FD results for different values of antennas plotted as baseline (blue lines). Solid/dashed lines correspond to non-linear/linear amplifier model.}
    \label{fig:N_evo_FD_comparative}
\end{figure}

We now aim to check whether the previous behaviour is consistent in different setups, in particular considering higher supplier power, more transmitters and users. This is illustrated in Fig. \ref{fig:N_evo_FD_comparative}, where we show the performance of both hybrid and \gls{DMA} topologies, compared against different \gls{FD} topologies (blue lines). Besides, we compare both amplifier models: linear and non/linear. We observe that the amplifier model is a game changer, considerably shifting the drawn conclusions (see, e.g., hybrid vs \gls{FD} performance). Hence, one should be cautious when generalizing these conclusions to other scenarios and setups. Regarding the \gls{DMA} vs. \gls{FD} comparison, the former outperforms (in terms of energy efficiency) the latter given enough radiating elements per waveguide, and always yields better results than the phased array-based solution. However, we notice that the gain of the \gls{DMA} w.r.t. hybrid arrays for the non-linear amplifier model is less pronounced than in Fig. \ref{fig:N_evo_ee_mse}, where $P_g^\text{max}$ is lower, which leads us to analyze the impact of the maximum supplied power. 

\begin{figure}[t]
    \centering
    \includegraphics[width=0.8\linewidth,trim={0 0.4cm 0 0.4cm}]{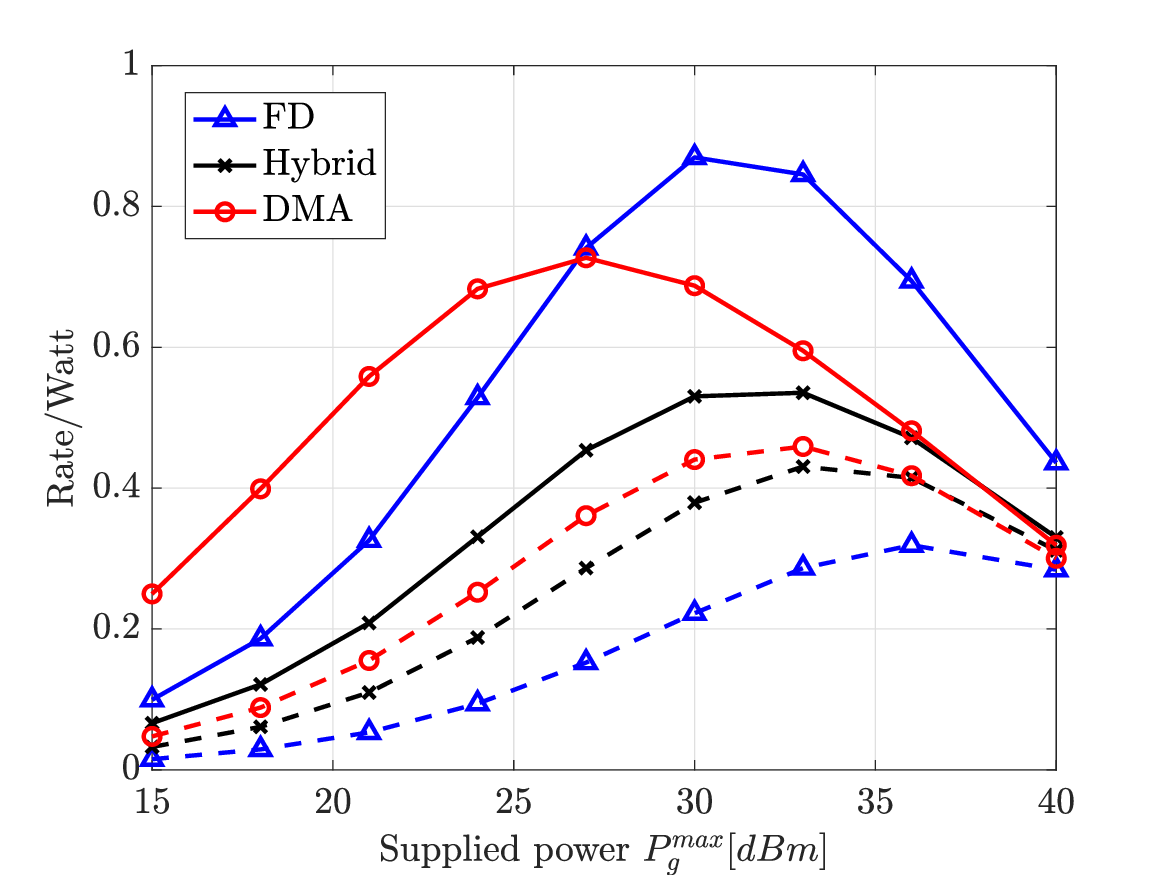}
    \caption{Energy efficiency vs $P_{g}^\text{max}$ for $N_\text{t} = 10$, $N = 120$, $M = 8$, spacing of $\lambda/2$ along the $x$ axis and $\lambda$ along the $z$ axis, non-linear amplifier model, and $\mat{E}[|(\mat{Y}_\text{ra})_{m,n}|^2]/\sigma_n^2 = 1$. Solid/dashed lines correspond to $P_\text{rf} = 40$ mW and $P_\text{rf} = 400$ mW.}
    \label{fig:Pgmax_evo}
\end{figure}

To that end, we plot in Fig. \ref{fig:Pgmax_evo} the evolution of the energy efficiency for different $P_g^\text{max}$. The same setup as in Fig. \ref{fig:N_evo_FD_comparative} is used, fixing $N/N_\text{t} = 12$ and increasing $\sigma_n^2$ in order to better appreciate the effect of $P_g^\text{max}$. We here keep the non-linear assumption but test two different power consumption values for the \gls{RF} chain: $40$ mW as in Table \ref{tab:SimParaBaseline} and $400$ mW as indicated in \cite{Auer2011} for a typical LTE picocell sector. In general, we observe the same trend as in other related works, where the energy efficiency increases with the supplied power up to some point, from which the trend reverts. Interestingly, we see that the \gls{DMA} architecture is more efficient with less available power, but cannot benefit from a larger $P_g^\text{max}$ as much as array-based solutions. This may be justified by how the different radiating structures work. In a phased array, regardless the phase shift applied, the available power can be equally split between the different subarrays. However, the radiating elements of the \gls{DMA} have a Lorentzian constraint response in isolation \cite{Smith2017, Williams2022}, which means that amplitude and phase shift are not independent. Another interesting remark is that, again, changing the power consumption values 
has a considerable impact, penalizing the \gls{FD} topology in this case. 

\begin{figure}[t]
    \centering
    \includegraphics[width=0.8\linewidth,trim={0 0.4cm 0 0.4cm}]{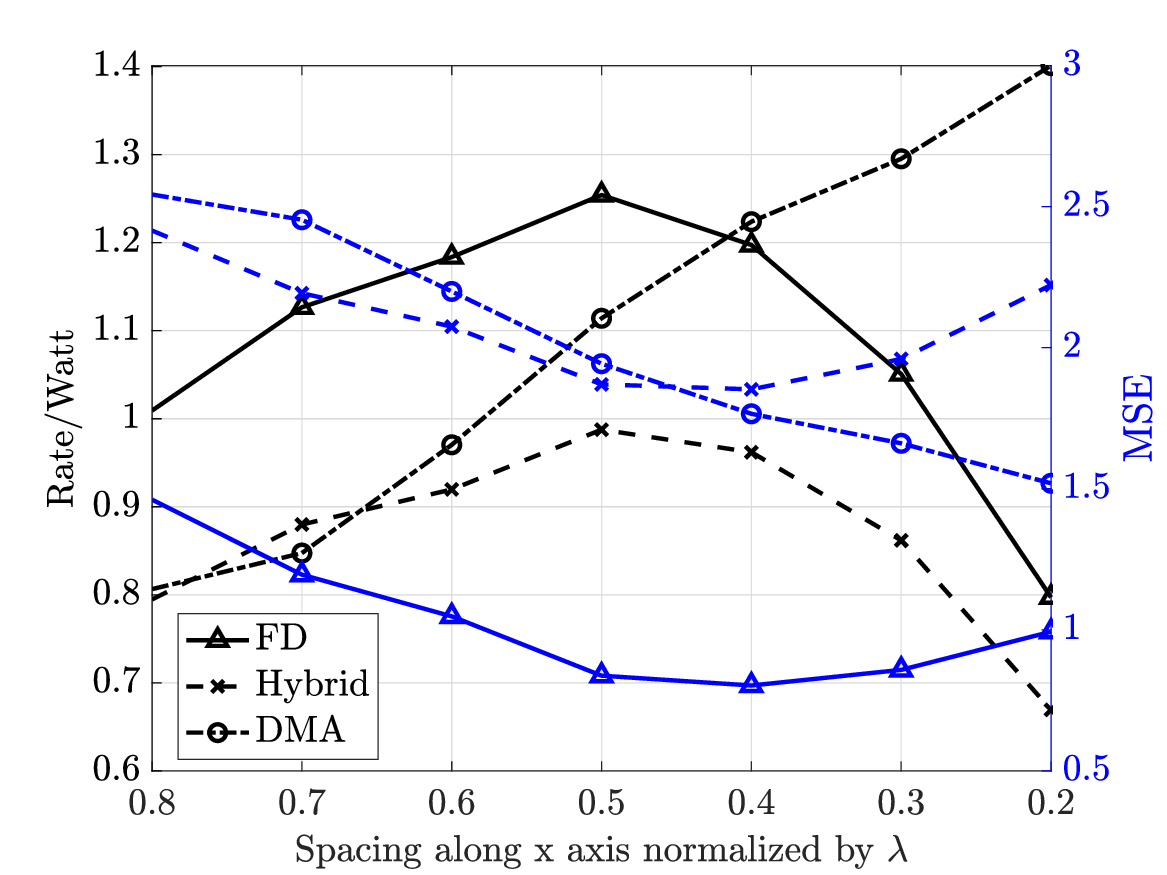}
    \caption{Energy efficiency and MSE achieved by the three topologies for different antenna spacings and fixed physical aperture. $N_\text{t} = 6$, $M = 5$, $P_g^\text{max} = 30$ dBm, and spacing of $3\lambda/4$ along the $z$ axis. $N/N_\text{t}$ ranges from 5 to 20, according to the spacing. Non-linear amplifiers assumed.}
    \label{fig:Spacing_evo}
\end{figure}

Finally, we explore the impact of densifying the array aperture by reducing the inter-antenna spacing. With this in mind, we represent in Fig. \ref{fig:Spacing_evo} both the energy efficiency and the achieved \gls{MSE} as the spacing along the $x$ axis is reduced. The spacing along the $z$ direction is fixed to $3\lambda/4$, to avoid overlapping of the waveguides in the \gls{DMA} (which have a width of $a = 0.73\lambda$). Hence, we keep the aperture constant, and start reducing the spacing, adding more antenna elements. Both \gls{FD} and hybrid arrays achieve the maximum \gls{MSE} performance for a spacing around $0.4\lambda$, a well-known result in wireless. Continuing to reduce the spacing leads to a worse (higher) \gls{MSE} despite the increase in number of antennas. In turn, the \gls{DMA} is able to take advantage of every antenna element, regardless the spacing. This is related to the insertion losses generated by the mutual coupling. As we reduce the spacing, the off-diagonal elements of $\mat{Y}_\text{aa}$ in the arrays become more relevant, and hence the reflection coefficient at the antenna interface increases (the norm in the difference $\mat{Y}_\text{aa} - Y_\text{g}\mat{I}_N$ increases), leading to less radiated power. For the \gls{DMA}, as $\mat{Y}_\text{p}$ depends on $\mat{Y}_\text{s}$ (the tunable admittances), we can configure the structure not only to achieve the desired radiation pattern but also to minimize the reflection coefficient. Naturally, diminishing returns will eventually be observed in terms of both \gls{MSE} and energy efficiency, as lower spacing implies more correlated channels and, at the end, the degrees of freedom are constrained by the physical aperture. 

\section{Conclusions}
\label{sec:Conclusions}
This work presents the building blocks for a thorough energy efficiency analysis of \gls{MIMO} systems, considering conventional arrays and \glspl{DMA} under the same/unified communications model and accounting for physical phenomena such as coupling and insertion losses in a natural way from a signal processing perspective, as exemplified in the derived transmit Wiener filtering solutions. 

\Glspl{DMA} emerge as very promising and more efficient alternatives to massive arrays, specially if: \textit{i)} low transmission/supplied power is available, \textit{ii)} strong coupling is expected, and \textit{iii)} a larger number of antennas need to be packed in a limited aperture. However, as shown throughout the paper, theoretical energy efficiency analyses are also rather sensitive to modeling issues, and small changes in the components consumption values may significantly shift the results. Caution is therefore advised with drawn conclusions, particularly those based on absolute results. Instead, it seems more sensible to analyze trends such as, e.g., the impact of low available power or the most consuming components, increasing numbers of antennas, or antenna spacing. 
Additional analyses, diving further into the impact of power consumption models, more ellaborated precoders and different frequencies, are intended for future works.

\bibliographystyle{IEEEtran}
\bibliography{references.bib}

\end{document}